\documentclass[twocolumn]{aa} 
\AddToHook{begindocument/before}{\usepackage{hyperref}} 

\usepackage[varg]{txfonts}
\usepackage{graphicx}
\usepackage{CJK}

\authorrunning{Li et al}
\titlerunning{A fast-rotating blue straggler star in the tidal tail of the open cluster NGC 752}

\begin{document} 
\begin{CJK}{UTF8}{gbsn}
%
%
%
\title{A fast-rotating blue straggler star in the tidal tail of the open cluster NGC 752 }


\author{Chunyan Li(李春燕)\inst{1,2}
          \and
          Jing Zhong(钟靖)\inst{1}
          \and
          Songmei Qin(秦松梅)\inst{1,2}
          \and
          Dengkai Jiang(姜登凯)\inst{2,3,4}
          \and
          Xingmei Shan(单星美)\inst{5}
          \and
          Li Chen(陈力)\inst{1,2}}

\institute{ Key Laboratory for Research in Galaxies and Cosmology, Shanghai Astronomical Observatory, Chinese Academy of Sciences, 80 Nandan Road, Shanghai 200030, China \\
        \and
        School of Astronomy and Space Science, University of Chinese Academy of Sciences, No. 19A, Yuquan Road, Beijing 100049, China. \\
        \and
        Yunnan Observatories, Chinese Academy of Sciences, 396 Yangfangwang, Guandu District, Kunming 650216, China \\
        \and 
        International Centre of Supernovae, Yunnan Key Laboratory, Kunming 650216, China \\
        \and
        Shanghai Science and Technology Museum, No. 2000 Century Avenue, Pudong New Area, Shanghai 200127, China \\
        {\it e-mail: lichunyan@shao.ac.cn,chenli@shao.ac.cn} }

\date{Received ???, ????; accepted ???, ????}

 \abstract
   {NGC 752 is a famous Galactic open cluster of intermediate age. In recent works, a very long and asymmetric tail was newly revealed. A blue straggler star (BSS) at the periphery of the tidal tail of the cluster has been identified subsequently.}
   {We aim to perform a detailed analysis of the newly detected BSS based on the available comprehensive spectroscopic and photometric data. We also explored this BSS's possible formation pathway and age limitation based on the collected spectroscopic and photometric data.}
   {We estimated the projected rotational velocity $v\ \mathrm{sin}i$ and the mass of the BSS from the Large Sky Area Multi-Object Fiber Spectroscopic Telescope low-resolution spectra and multiband photometric data from various catalogs, respectively.}
   {The newly discovered BSS is confirmed as a genuine member of NGC 752. The lack of ultraviolet excess in the SED and no significant variations in the light curve imply that this BSS is likely a single star ($mass=1.86^{+3.62}_{-0.94}\ M_{\odot}$) formed through stellar mergers. The fast rotation velocity ($v\ \mathrm{sin}i=206.9\pm4.9$~km $\rm s^{-1}$) of the BSS may provide constraints on its age (less than a hundred million years), but more formation details require further investigation.}
   {}
   
\maketitle
%
\section{Introduction \label{sect:intro}}
Blue straggler stars (BSSs) are typically characterized as a population of stars that are bluer and brighter than the main sequence turn-off (MSTO) stars. 
They are observed as an extension of the main sequence positioned above the MSTO in the color-magnitude diagram (CMD). Hence, BSSs are thought to be more massive than MSTO stars \citep{1997ApJ...489L..59S, 1998ApJ...507..818G, 2005ApJ...632..894D, 2014ApJ...783...34F}, which implies that BSSs are the products of main sequence stars that gained extra mass.

BSSs were first reported by \cite{Sandage_1953} in the CMD of the globular cluster M3. In the Milky Way, BSSs are not only commonly identified in globular clusters (GCs) \citep{Piotto_2004,2008ApJ...678..564L} or dwarf spheroidal galaxies \citep{2012MNRAS.421..960S}, but also exist in the intermediate-age or old open clusters (OCs) \citep{Ahumada_2007, Rain_2021, Jadahav2021, 2023A&A...672A..81L}. 

There are multiple formation scenarios for creating BSSs: (i) direct stellar collisions during dynamical encounters \citep{1976ApL....17...87H, 1989AJ.....98..217L}; (ii) mass transfer from binary companions \citep{1964MNRAS.128..147M}; (iii) mergers of two main sequence stars \citep{2001ApJ...552..664N, Perets_Fabrycky_2009}. Although the relationship between formation environment and the mechanism is still not completely understood \citep{2007ApJ...661..210L}, collisional BSSs are expected to be produced more likely in the high-density environment and have been identified in a few GCs \citep{2009Natur.462.1028F, 2013ApJ...778..135D, 2019ApJ...876...87B, 2022ApJ...941...69C}. However, BSSs formed from the evolution of binaries are also widely found in these GCs. Presently, the main leading BSS formation scenarios involve mass transfer processes between binary companions, possibly up to the complete coalescence of the binary system, or mergers of stars induced by collisions \citep{2015ASSL..413...99F}. It is plausible that these formation mechanisms could be at work simultaneously in a given GC or OC.
\cite{2006ApJ...646.1160A} quantified the angular momentum loss induced by magnetic stellar winds in tidally synchronized binaries on the main sequence and suggested that this mechanism is responsible for at least one-third of the BSSs in OCs older than 1 Gyr. It was also proposed in recent studies that BSS formed from the merger of main sequence stars previously in a hierarchical triple system as a result of the eccentric Kozai--Lidov mechanism \citep{Perets_Fabrycky_2009, Naoz_2014}, which has a significant role in BSS formation in OCs. Indeed, plenty of evidence supports the notion that multiple formation scenarios contribute to creating BSSs in open clusters \citep{Jadahav2021,2021ApJ...908..229L}.

NGC 752 is an open cluster with intermediate age of 1-1.5 Gyr \citep{2018ApJ...862...33A, 2020A&A...640A...1C, 2022MNRAS.514.3579B}. 
After comparing the initial mass and the present-day mass of NGC 752, it is found that this cluster is experiencing a disintegration process since it has lost nearly 95.2\%-98.5\% of its initial mass due to stellar evolution and tidal interactions \citet{2021MNRAS.505.1607B}.
NGC 752 is centered at ($l = 136.959 \degr$, $b = -23.289 \degr$) and located about 483 pc away from the Sun, exhibiting a very small extinction \citep{2020A&A...640A...1C,2021MNRAS.504..356D}. In the earlier works, the cluster radius of the cluster is estimated as 84 arcmin \citep{2013A&A...558A..53K}, and the mean half number radius is about 28.2 arcmin \citep{Cantat-Gaudin_2020, 2020A&A...640A...1C, 2021MNRAS.504..356D}. 

However, with the help of {\it Gaia} data \citep{gaia18, 2021A&A...649A...1G}, more and more studies have revealed the existence of outer halo region or extended tidal tails \citep{2019A&A...624A..34Z,2022AJ....164...54Z,2020ApJ...889...99Z, 2022RAA....22e5022B, 2022ApJ...931..156P}, leading to a significant increase in the size of OCs. Using {\it Gaia} EDR3 data \citep{2021A&A...649A...1G},
\cite{2021MNRAS.505.1607B} extended the searching area for cluster member stars to a $5 \degr$ radius around the center of NGC 752 and identified its tidal tails spanning up to 35 pc on both sides of the cluster center.
Furthermore, \cite{2022MNRAS.514.3579B} investigated a much wider region around NGC 752 and revisited its tidal tails.
It is found that the tidal tails of NGC 752 are considerably longer than what was previously claimed in \cite{2021MNRAS.505.1607B} and span over 260 pc in the sky (equivalent to $37 \degr$ at the distance of NGC 752).

NGC 752 was first reported to host a BSS in the central area by \cite{1994PASP..106..281D}. But this BSS (hereafter named BS\_D94) was absent from the BSS catalog of \cite{Ahumada_2007}, which was referred to as the comprehensive catalog of BSSs in the Galactic open clusters up to 2005. This inconsistency may be attributed to different cluster membership determination results by various works, leading to different authentication results for BSS. In the {\it Gaia} era, \cite{Rain_2021} identified 897 BSSs in 111 OCs using cluster member stars provided by \cite{Cantat-Gaudin_2020}. Similarly, \cite{Jadahav2021} produced a catalog of 868 BSSs in 228 OCs with the membership information from \cite{2020A&A...640A...1C}. The two catalogs can be considered as an update to the catalog of \cite{Ahumada_2007}. However, the BS\_D94 was not included in either catalog because it was not regarded as a member star of NGC 752 by \cite{Cantat-Gaudin_2020, 2020A&A...640A...1C}.

To identify more BSSs in a larger OC region, \cite{2023A&A...672A..81L} extended the searching area and re-performed the membership determination of OCs, which was accomplished based on astrometric data from {\it Gaia} DR3 \citep{2023A&A...674A...1G} by the pyUPMASK\footnote{a python version of the `Unsupervised Photometric Membership Assignment in Stellar Clusters' algorithm (UPMASK)} algorithm \citep{2014A&A...561A..57K, 2021A&A...650A.109P}. Finally, \cite{2023A&A...672A..81L} identified 138 new BSSs in 50 open clusters, including a BSS (hereafter BS\_Li23) in NGC 752. In particular, the BS\_Li23 was discovered at the farther end of the tidal tail with an angular distance extending over $8 \degr$ from the center of NGC 752.

According to the results of \cite{2023A&A...672A..81L}, we found that approximately 91\% of BSSs are distributed within three times half number radius ($R_{h}$) of their host OCs. However, the location of BS\_Li23 in NGC 752 exceeds three times $R_{h}$, which may indicate that the formation process of BS\_Li23 differs from most BSSs. We thus have conducted a detailed analysis of the observational properties of BS\_Li23 and further probe the possible formation way of BSSs in a low-density environment of open clusters.

In this work, we collected the optical spectroscopy, multiband photometry, and time-domain photometry observational information and then performed a detailed analysis of the BSS in NGC 752 reported by \cite{2023A&A...672A..81L}.
In Sect. \ref{sect:d&m}, we analyzed the available spectroscopic and photometric data of BS\_Li23 and extract properties of this BSS. We discussed the probable formation pathway and the age limitation of the BSS in Sect. \ref{sect:pathway}. In Sect. \ref{sect:D94}, we reevaluated the membership of BS\_D94 based on {\it Gaia} DR3 data. Finally, we give a summary in Sect. \ref{sect:sum}.
\section{Properties of the BS\_Li23 \label{sect:d&m}}
\begin{figure}
   \centering
   \includegraphics[width=0.75\hsize]{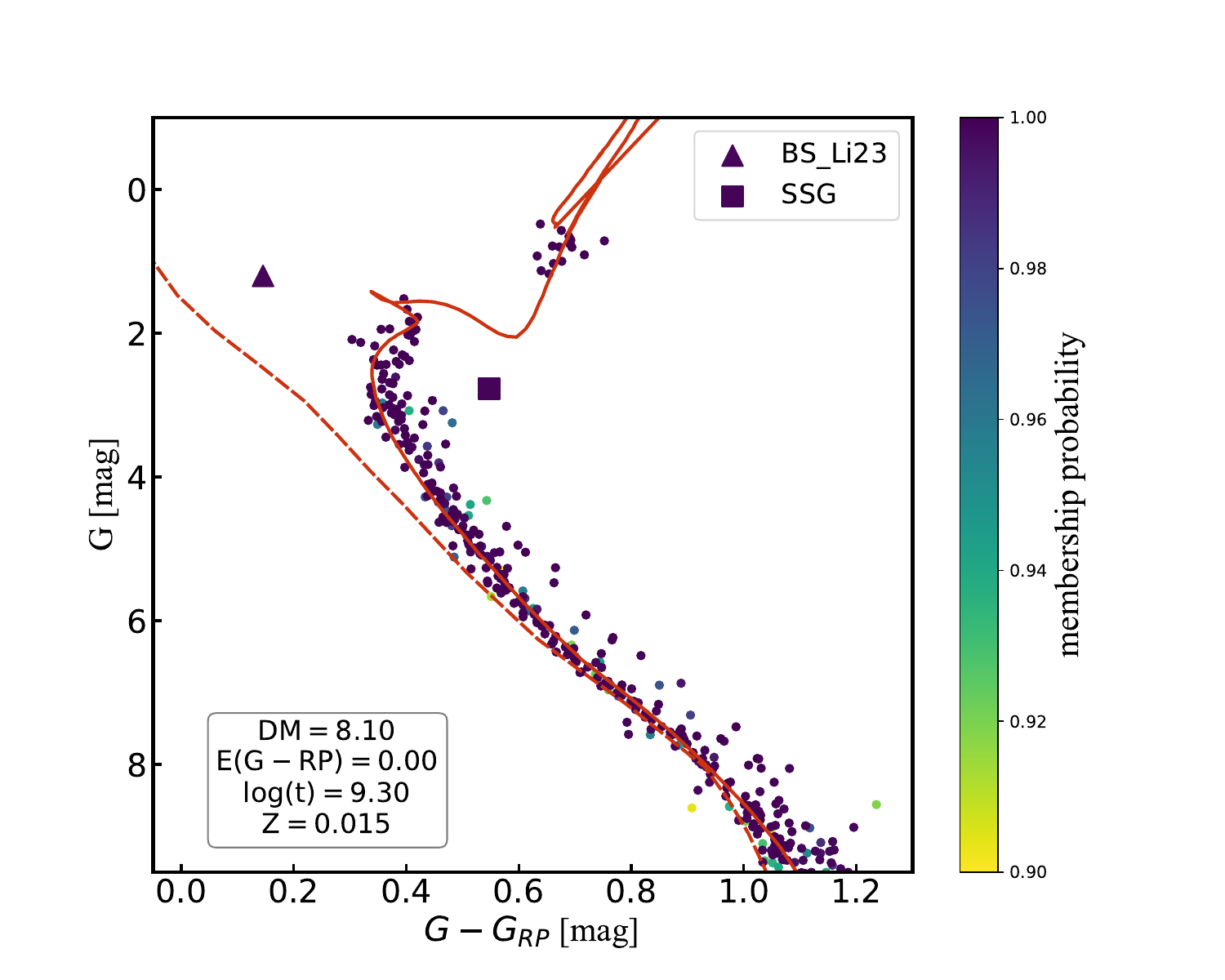}
      \caption{Differential reddening corrected CMD of NGC 752. The triangle and square symbols represent the BSS of NGC 752 reported by \cite{2023A&A...672A..81L} and an SSG candidate, respectively, while the dots denote cluster member stars. The red solid line represents the best-fitting isochrone, with parameters provided by \cite{2023A&A...672A..81L}. The red dashed line corresponds to the ZAMS. The color bar indicates the membership probability.}
         \label{fig: paper2_bs_CMD}
\end{figure}

In this study, all members of NGC 752 are from the catalog of \cite{2023A&A...672A..81L}. In total, 404 member stars with membership probabilities ($P_{m}$) no less than 0.9 were identified in NGC 752 by \cite{2023A&A...672A..81L}, and these stars are regarded as NGC 752 member samples in this work.
Considering that the differential reddening may introduce some dispersion on the CMD, especially near MSTO, which can affect the selection of BSS, we performed differential reddening (DR) correction to the CMD of NGC752 referring to the method of \cite{Milone_2012}.
As shown in Fig. \ref{fig: paper2_bs_CMD}, the DR corrected CMD of NGC 752 presents a clear main sequence and a main sequence turn-off branch while a member star is located within the region delimited by the best-fitting isochrone and Zero Age Main Sequence (ZAMS). We regarded this member star as a genuine BSS since it is located in the definition region of the typical BSSs in the CMD \citep{Ahumada_1995, Ahumada_2007, Rain_2021}. 
In this work, we named this BSS identified in NGC 752 as BS\_Li23.

It is worth noting that in Fig. \ref{fig: paper2_bs_CMD} there is a star at the reddest level of MSTO and fainter than the base of the sub-giant branch, marked as a filled square. From its unique location on the CMD, this star might be a potential sub-subgiant star(SSG). In the CMDs of star clusters, SSGs reside to the red of the main sequence but fainter than the subgiant branch, and are also known as `red stragglers' \citep{1998A&A...339..431B, 2003AJ....125..246M}. The origins of SSGs are not completely understood, as they are situated far from the predicted locus of single stars based on stellar evolution theory \citep{2015AJ....150...97G, 2017ApJ...840...66G,2017ApJ...840...67L,2017ApJ...842....1G}.This star may be the first identified SSG in NGC 752, well worth further detailed investigations on its characters and formation pathway in the future work.

\begin{table*}
\caption{Parameters of BS\_Li23}
\label{table:1}      
\centering          
\begin{tabular}{llrc}     
\hline\hline       
parameter& unit & value & source  \\
\hline 
RA (J2016) & deg & 39.0595 & {\it Gaia} DR3 \\
DEC (J2016) & deg & 34.7482  & {\it Gaia} DR3 \\
$\mu_{\alpha^{*}}$ & mas $\rm yr^{-1}$  & $10.24\pm 0.03$  & {\it Gaia} DR3 \\
$\mu_{\delta}$ & mas $\rm yr^{-1}$ & $-11.73\pm 0.02$  & {\it Gaia} DR3 \\
Plx & mas & $2.28\pm0.02$& {\it Gaia} DR3\\
RV & km $\rm s^{-1}$ & $4.66\pm 1.82$& {\it Gaia} DR3\\
$v\ \mathrm{sin}i _{\rm esphs}$  & km $\rm s^{-1}$ & $179.18\pm 3.71$& {\it Gaia} DR3\\
G & mag &9.422 & {\it Gaia} DR3\\
$\rm G_{\rm RVS}$ & mag & $9.162\pm0.005$ & {\it Gaia} DR3\\
$T_{\rm eff}$ & K & $8310\pm39$ &LAMOST DR9 \\
log$(g)$ & dex & $3.96\pm0.05$ &LAMOST DR9 \\ 
$\rm [Fe/H]$ & dex & $-0.095\pm0.033$ &LAMOST DR9 \\ 
$v\ \mathrm{sin}i_{H_{\alpha}}$ & km $\rm s^{-1}$& $210.3^{+2.5}_{-2.6}$& this work\\
$v\ \mathrm{sin}i_{H_{\beta}}$ & km $\rm s^{-1}$& $200.0^{+3.0}_{-2.6}$ & this work\\
$v\ \mathrm{sin}i_{H_{\gamma}}$ & km $\rm s^{-1}$& $210.5^{+1.5}_{-2.5}$ & this work\\
Mass & $\rm M_{\odot}$& $1.86^{+3.62}_{-0.94}$ & this work\\
\hline                  
\end{tabular}
\end{table*}

\subsection{Fundamental parameters\label{sect:fund}}
\begin{figure}
  \centering
  \includegraphics[width=0.75\hsize]{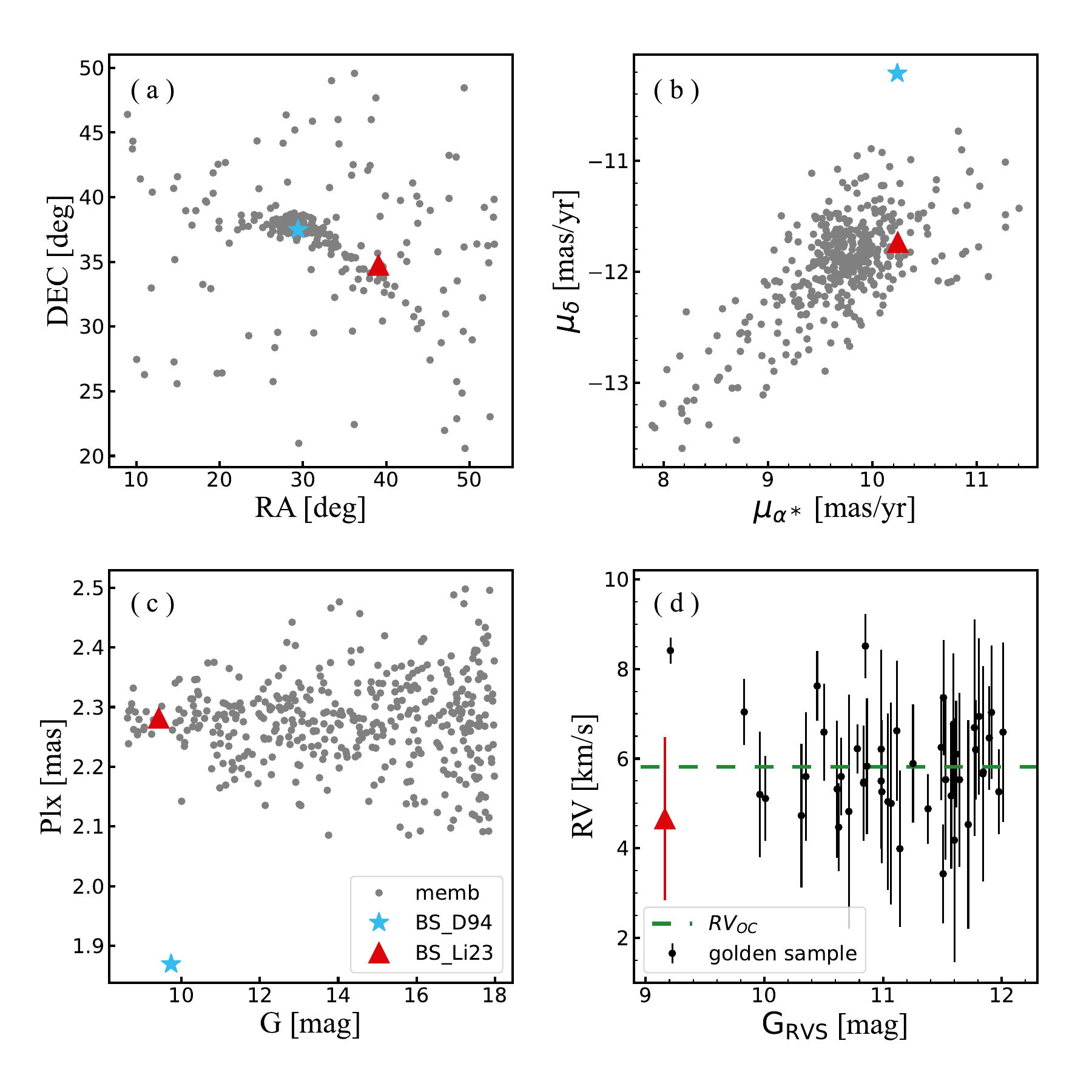}
      \caption{Spatial distribution, proper motion distribution, magnitude-parallax diagram, and radial velocity distribution of NGC 752 member stars identified in \cite{2023A&A...672A..81L} and a BSS reported by \cite{1994PASP..106..281D}. (a) to (c) panels: The grey dots represent the cluster members of NGC 752 with $P_{m} \geq 0.9$; For comparison, the star categorized as a BSS of NGC 752 by \cite{1994PASP..106..281D} is symbolized by a blue pentagram, labeled `BS\_D94'. (d) panel: Black dots mean member stars meeting the criteria of the golden sample described in Sect. \ref{sect:fund}; The radial velocity and corresponding error is from {\it Gaia} DR3; The green dashed line presents the mean RV value of the golden sample. The BSS reported by \cite{2023A&A...672A..81L} is marked as a red triangle labeled `BS\_Li23' in all panels.}
         \label{fig: paper2_members4d}
\end{figure}

As shown in panel (a) of Fig. \ref{fig: paper2_members4d}, BS\_Li23 is situated about $8.3\degr$ away from the cluster center ($RA=29.404\degr$, $DEC=37.7569\degr$); 
nevertheless, the proper motion and parallax parameters of BS\_Li23 are in accordance with other genuine NGC 752 member stars, as can be seen in panels (b) and (c) of Fig. \ref{fig: paper2_members4d}.

The fundamental parameters for BS\_Li23 provided by {\it Gaia} DR3 are listed in Table \ref{table:1}.
We also presented the radial velocity (RV) distribution of NGC 752 member stars in panel (d) of Fig. \ref{fig: paper2_members4d}.
As reported in \cite{2023A&A...674A...5K}, at $G_{\rm RVS} = 12$ mag, the median formal precision of radial velocities can reach 1.3 km $\rm s^{-1}$, and the exclusion of sources with low signal to noise ratio spectra (SNR less than 10) effectively eliminates a notable proportion of outliers. 
We employed member stars with $G_{\rm RVS} \leq 12$, $\rm SNR>10$ to estimate the average RV of NGC 752. After further removing four outliers, 45 main sequence stars are retained as a golden sample, and the mean RV of these stars is obtained to be $5.81$~km $\rm s^{-1}$ with the velocity dispersion of $1.06$~km $\rm s^{-1}$.

Moreover, there are 18 golden sample stars included in the Large Sky Area Multi-Object Fiber Spectroscopic Telescope (LAMOST) Medium-Resolution Spectroscopic (MRS) Survey Parameter Catalog\footnote{\url{http://www.lamost.org/dr10/v1.0/catalogue}}. The LAMOST, also named the Guo Shou Jing Telescope, is a 4-meter quasi-meridian reflective Schmidt telescope with 4000 fibers \citep{2012RAA....12.1197C}, which was equipped with a medium-resolution spectrograph (R$\sim$7500) with two arms covering the wavelength ranges of 495-535 nm and 630-680 nm, respectively \citep{2020arXiv200507210L}.
The average precision of RV obtained from LAMOST MRS spectra reaches 1.0~km $\rm s^{-1}$ \citep{2019RAA....19...75L,2020ApJS..251...15Z}. The mean LAMOST RV of NGC 752 is 6.17~km $\rm s^{-1}$ with a dispersion of 1.35~km $\rm s^{-1}$, which is similar to the result of {\it Gaia} RV.

It is noticed that the {\it Gaia} DR3 radial velocity of BS\_Li23 is $4.66\pm1.82$ km $\rm s^{-1}$, showing good consistency with the radial velocity characteristic of NGC 752. Supplemented by {\it Gaia} DR3 radial velocity data, we thus re-confirmed that BS\_Li23 is a genuine member BSS of NGC 752.

\subsection{Rotational velocity \label{sect:spect}}
\begin{figure}
   \centering
   \includegraphics[width=0.75\hsize]{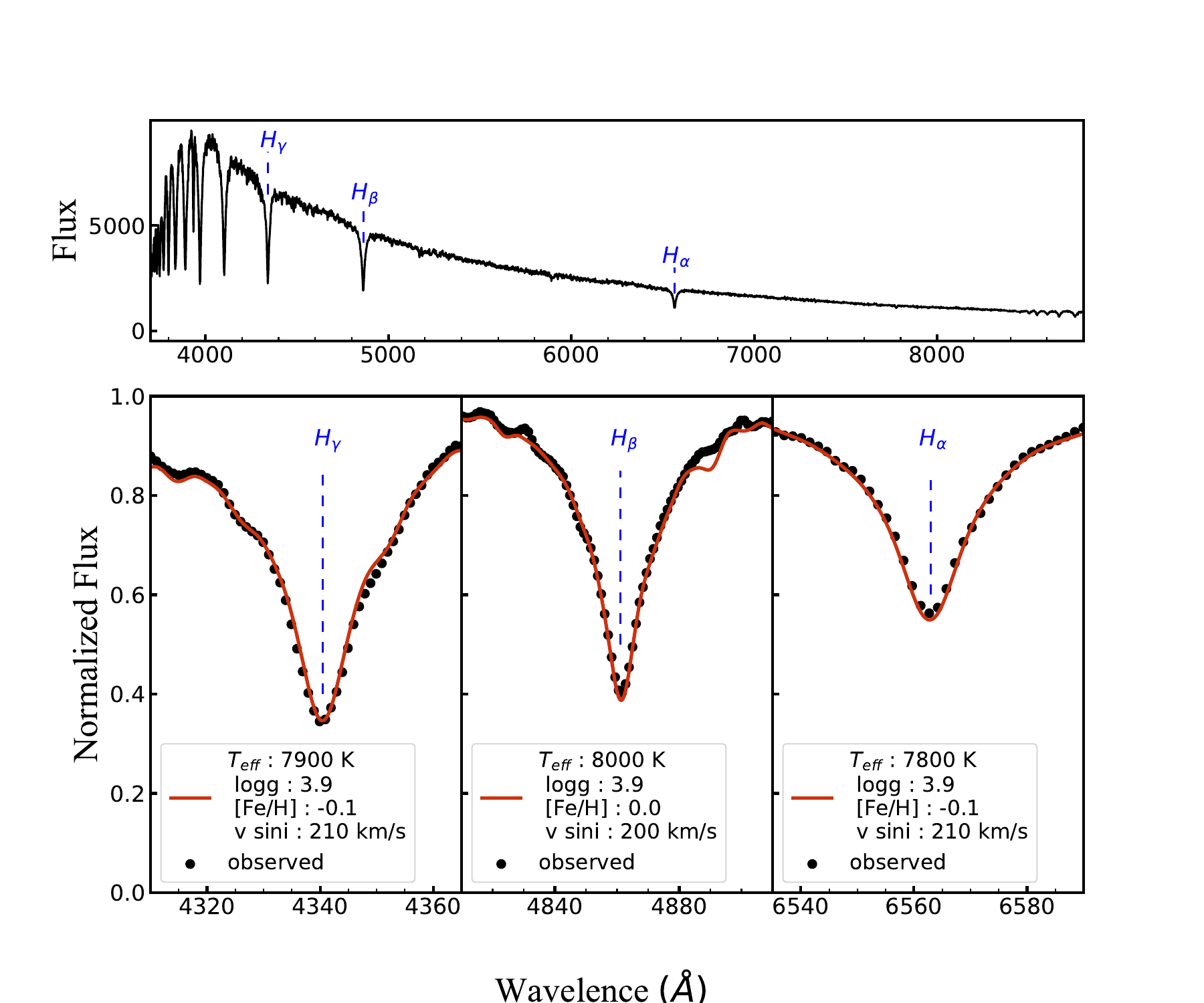} 
      \caption{Top panel: The low-resolution LAMOST spectrum of BS\_Li23, showing the Balmer ($H_{\alpha}$, $H_{\beta}$, and $H_{\gamma}$ absorption lines). Bottom panel: $H_{\gamma}$, $H_{\beta}$, and $H_{\alpha}$ lines along with their best-fitting models (red solid lines). Best-fitting parameters are shown in each panel.}
         \label{fig: paper2_LM_LRS}
\end{figure}

The projected rotational velocity ($v\ \mathrm{sin}i$) of BS\_Li23 is provided by the Extended Stellar Parametrizer for Hot Stars (ESP-HS) of {\it Gaia} DR3, and its value is listed in Table \ref{table:1}. 
However, considering the limitation of the spectra wavelength (845-872 nm), ESP-HS values suffer from the poor $v\ \mathrm{sin}i$-related information for OBA stars in this wavelength domain \citep{2023A&A...674A..28F}. 
Besides {\it Gaia} DR3 data, the only spectrum data we can find for BS\_Li23 is from the low-resolution spectra (LRS) of LAMOST Data Release 9 (DR9), whose designation name is J023614.26+344453.6 with an SNR of 60 at the effective wavelengths of Sloan DSS $g$ filter \citep{1996AJ....111.1748F, 2006AN....327..821T}.
The wavelength range of the low-resolution spectra spans from 370 to 900 nm, and the spectral resolution is 1800 \citep{2012RAA....12..723Z}.
The LAMOST spectrum of BS\_Li23, as shown in the top panel of Fig. \ref{fig: paper2_LM_LRS}, indicates it is an A-type star.
The basic stellar parameters of BS\_Li23 are provided in the LAMOST LRS Stellar Parameter Catalog of A, F, G, and K Stars\footnote{\url{http://www.lamost.org/dr9/v2.0/catalogue}}. 
In the catalog, the stellar spectral type of BS\_Li23 is classed as A5, and the atmosphere parameters of the BS is $T_{\rm eff}=8310\pm39$ K, log$(g)=3.96\pm0.05$, [Fe/H]=$-0.095\pm0.033$, which were determined by the LAMOST stellar parameter pipeline (LASP).
However, this catalog does not provide the projected rotational velocity of stars.

To estimate the $v\ \mathrm{sin}i$ of BS\_Li23 with the LAMOST data, we employed $\chi^{2}$ minimization method by comparing the observed LAMOST LRS to a grid of synthetic spectra for different values of $v\ \mathrm{sin}i$. Considering the spectral resolution (R$\sim$1800) of LAMOST spectrum is low and the spectral type of BS\_Li23 is A type, we choose to use the Balmer lines ($H_{\alpha}$, $H_{\beta}$, and $H_{\gamma}$) rather than other weak metal lines to determine the $v\ \mathrm{sin}i$ parameter.
These high-resolution synthetic spectra were generated using the iSpec code \citep{2014A&A...569A.111B, 2019MNRAS.486.2075B}, an open-source framework for spectral analysis that integrates ATLAS9 model atmospheres \citep{2005MSAIS...8...14K} with the radial transfer codes SYNTHE \citep{1993KurCD..13.....K, 2004MSAIS...5...93S}.
Referring to the parameters from the LAMOST DR9 catalog (listed in Table \ref{table:1}) for BS\_Li23, the synthetic spectra were derived with $T_{\rm eff}$ range from 7500~K to 9500~K (in steps of 100~K), log$(g)$ from 3.9~dex to 4.0~dex (in steps of 0.1~dex) and [Fe/H] from -0.2~dex to 0.2~dex (in steps of 0.1~dex).
The microturbulence velocities of main sequence early-type stars were about 0-3~km $\rm s^{-1}$ in previous works \citep{1971A&A....11..387L, 1972ApJ...177..115M, 1973ApJ...179..209M,2011A&A...528A..44M}.
Hence, we assumed a microturbulent velocity of 2~km $\rm s^{-1}$ for BS\_Li23 when generating the synthetic spectra.
All the synthetic spectra have been downgraded to the resolution from 80000 to 1800 to match the LAMOST low-resolution spectra.
Then we convolved the synthetic spectra with different rotational kernels \citep{2005oasp.book.....G} from $v\ \mathrm{sin}i$ = 5 km $\rm s^{-1}$ to 300 km $\rm s^{-1}$ (in steps of 2~km $\rm s^{-1}$).
We further used a Monte Carlo method \citep{2013PASP..125..306F} to estimate the uncertainty in $v\ \mathrm{sin}i$ around the grid values.
Finally, the synthetic spectra that minimized the $\chi^{2}$ are shown in the bottom panel of Fig. \ref{fig: paper2_LM_LRS}, along with the parameters of these synthetic spectra and corresponding best-matched $v\ \mathrm{sin}i$.
The estimated values of $v\ \mathrm{sin}i$ from the fitting results of $H_{\alpha}$, $H_{\beta}$, and $H_{\gamma}$ lines are similar ($v\ \mathrm{sin}i_{H_{\alpha}}=210.3^{+2.5}_{-2.6}$~km $\rm s^{-1}$, $v\ \mathrm{sin}i_{H_{\beta}}=200.0^{+3.0}_{-2.6}$~km $\rm s^{-1}$, $v\ \mathrm{sin}i_{H_{\gamma}}=210.5^{+1.5}_{-2.5}$~km $\rm s^{-1}$).
The average value of $v\ \mathrm{sin}i$ obtained from the results of the three lines is about $206.9\pm 4.9$~km $\rm s^{-1}$.

\cite{2016A&A...594A..39F} mentioned that the $v\ \mathrm{sin}i$ determined from LAMOST LRS spectra can be considered reliable if the value exceeds 120~km $\rm s^{-1}$. As stated in a recent work by \cite{2024ApJS..271....4Z}, the LAMOST $v\ \mathrm{sin}i$ measurement precision is mainly affected by the SNR of spectra. They calculated the $v\ \mathrm{sin}i$ precision for 80108 stars with multiple LAMOST LRS observations and estimated that the intrinsic precision of $v\ \mathrm{sin}i$ is about $10$~km $\rm s^{-1}$ when $\rm SNR>50$. This precision is acceptable for BS\_Li23 with a large value of $v\ \mathrm{sin}i$ over $200$~km $\rm s^{-1}$. Compared our result with {\it Gaia} DR3 $v\ \mathrm{sin}i _{\rm esphs}$, the difference is about $30$~km $\rm s^{-1}$, corresponding to about 15\% relative deviation at the high rotational velocity of about $200$~km $\rm s^{-1}$. Since the projected rotational velocity parameter has not been provided in the LAMOST LRS Stellar Parameter Catalog, we lack information regarding the overall systematic differences between LAMOST and {\it Gaia} in $v\ \mathrm{sin}i$. 
On the other hand, \cite{2024ApJS..271....4Z} found that the $v\ \mathrm{sin}i$ they estimated from LAMOST LRS spectra generally agree with {\it Gaia} $v\ \mathrm{sin}i _{\rm esphs}$ with a scatter of $24.7$ km $\rm s^{-1}$ for 7013 common stars, which is also compatible to the difference in our work.

Therefore, we regard BS\_Li23 as a genuine fast-rotating BSS, given that it is relatively rare for BSSs to rotate faster than $50$~km $\rm s^{-1}$ in both GCs \citep{2023NatCo..14.2584F, 2023ApJ...956..124B} and OCs \citep{2009Natur.462.1032M, 2023AJ....166..154B}.

\subsection{Stellar mass \label{sect:sed}}
\begin{figure}
   \centering
   \includegraphics[width=0.75\hsize]{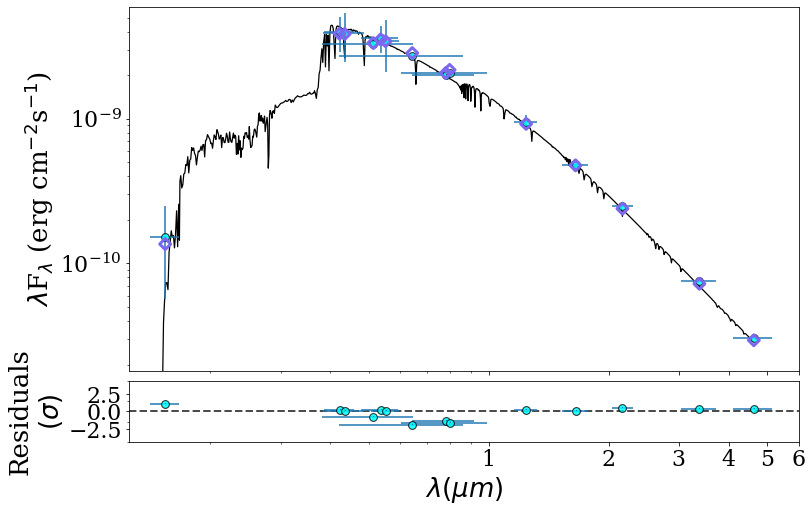}
      \caption{Best-fitting SED model for BS\_Li23. The black curve represents the best-fitting model, while the cyan pluses and circles denote the retrieved photometric measurements. The blue diamonds correspond to synthetic photometry.}
    \label{fig: paper2_SED}
\end{figure}

In our investigation of the blue straggler BS\_Li23, we used spectral energy distribution (SED) fitting to constrain its stellar mass.
The SED fitting was performed by a python package ARIADNE\footnote{\url{https://github.com/jvines/astroARIADNE}}, which aims to use Bayesian Model Averaging to incorporate the information from as many as six distinct atmospheric model grids to arrive at accurate and precise stellar parameters \citep{2022MNRAS.513.2719V}. We collected multiband photometric data from various catalogs to fit the SED, including GALEX (FUV), TYCHO ($B$ and $V$), {\it Gaia} DR3 ($G$, $G_{\rm BP}$, and $G_{\rm RP}$), {\it TESS} ($T$), 2MASS ($J$, $H$, and $K_{\rm s}$), and WISE ($W{\rm 1}$ and $W{\rm 2}$).
We crossmatched BS\_Li23 with other UV photometric catalogs but only obtained FUV band data from GALEX, which has been utilized in our SED fitting.
Then, we used the atmospheric parameters from LAMOST DR9, parallax from {\it Gaia} DR3, and an extinction value of $A_{v}=0$ obtained from \cite{2023A&A...672A..81L} isochrone fitting result as prior parameters for the SED fitting. 
Fig. \ref{fig: paper2_SED} shows the SED fitting result. A single-star model can accurately represent the SED of BS\_Li23. The best-fitting parameters with their 68\% confidence intervals are as follows: $T_{\rm eff}=8068^{+48.49}_{-83.55}$, log$(g)=4.05^{+0.23}_{-0.26}$, and $R=2.718^{+0.033}_{-0.038}\ R_{\odot}$.
We then calculated the stellar mass of BS\_Li23 as $1.86^{+3.62}_{-0.94}\ M_{\odot}$ based on the derived stellar gravity and radius.

\begin{figure}
   \centering
   \includegraphics[width=0.75\hsize]{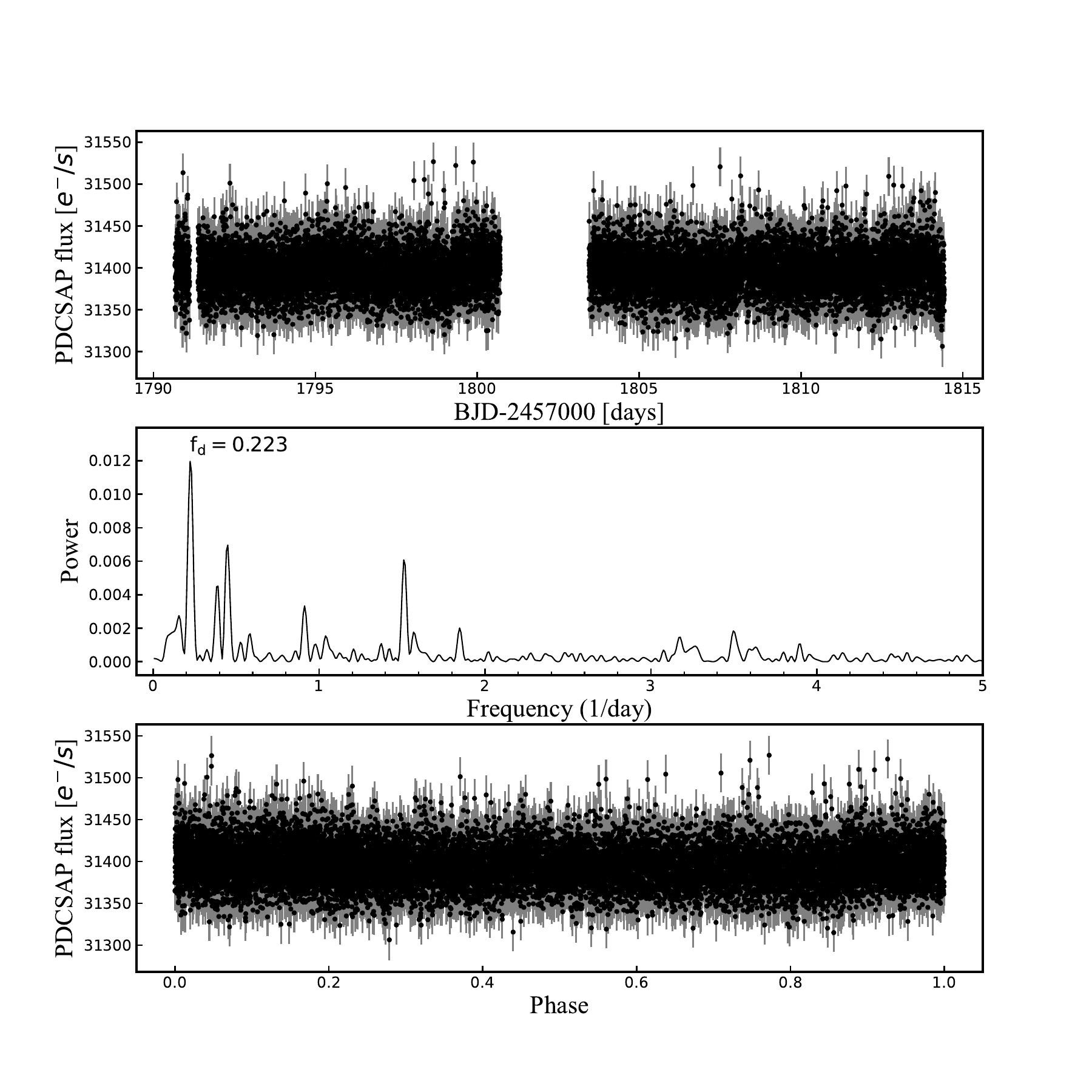}
      \caption{Top panel: TESS Pre-search Data Conditioning (PDC) corrected Simple Aperture Photometry (SAP) light curve. Middle panel: Lomb-Scargle results for BS\_Li23. Bottom panel: folded corrected Simple Aperture Photometry (SAP) light curve at the dominant period.}
         \label{fig: paper2_TESS}
\end{figure}

To explore the possible binarity property of BS\_Li23, we first checked the variability flag of this star in {\it Gaia} DR3, but the flag of `PHOT\_VARIABLE\_FLAG' is marked as `NOT\_AVAILABLE' for this star.
Then we analyzed the light curve from {\it TESS} \citep{2015JATIS1a4003R}, which has been collected by {\it TESS} sectors 18 and 58. The light curve for this star, processed by the {\it TESS} Science Processing Operations Center (SPOC, \cite{2016SPIE.9913E..3EJ}), is available on the Mikulski Archive for Space Telescopes (MAST) website\footnote{\url{https://archive.stsci.edu/missions-and-data/tess}} and shown in top panel of Fig. \ref{fig: paper2_TESS}. 
The PDCSAP\_FLUX light curve is a detrended flux time series and also corrects for the amount of flux captured by the photometric aperture and crowding from known nearby stars. Then, we performed the Lomb-Scargle algorithm \citep{lomb1976least, 1982ApJ...263..835S} to identify the potential period of the PDCSAP\_FLUX light curve. The middle panel of Fig. \ref{fig: paper2_TESS} shows the Lomb-Scargle Periodogram, revealing a dominant frequency of approximately 0.223~$\rm day^{-1}$. In the bottom panel of Fig. \ref{fig: paper2_TESS}, the light curve in the top panel is folded at this period. However, hardly any variations in the phase-folded light curve of BS\_Li23 could be detected. Therefore, BS\_Li23 is unlikely to be a binary.

\section{Possible formation pathways and age limitation of BS\_Li23 \label{sect:pathway}}
Several mechanisms were proposed for the formation of BSSs. Three major formation pathways of BSSs include (i) collisions during stellar dynamical encounters \citep{1989AJ.....98..217L}; (ii) mass transfer from a binary companion \citep{1964MNRAS.128..147M}; (iii) mergers of close binary systems, perhaps the inner binaries in triple star systems driven to a merger by Kozai cycles \citep{Perets_Fabrycky_2009}.

Star collisions are generally believed to produce BSS, which is more common in environments with high stellar density, such as globular clusters. Compared to globular clusters, the density of open clusters is much lower, and the simulation indicates that the probability of producing BSSs through collisions in open clusters is very low \citep{2013AJ....145....8G}. Furthermore, considering that BS\_Li23 is located on the tidal tail of the disintegrating open cluster NGC 752 \citep{2021MNRAS.505.1607B}, its surrounding stellar density is much lower than the average cluster density. Therefore, the probability of this star being produced due to stellar collisions could be ignored. 

Most BSSs are believed to have formed in open clusters through binary evolution \citep{2015ASSL..413...29M}. This mechanism will generate a BSS + white dwarf (WD) system or a single BSS. The existence of a white dwarf companion can be verified by examining whether there is an ultraviolet excess on the SED or flux variation in the light curve. 
As shown in Fig. \ref{fig: paper2_SED} and Fig. \ref{fig: paper2_TESS}, no ultraviolet excess appeared on the SED of BS\_Li23 (the ultraviolet data comes from the FUV band of GALEX), and no observable significant optical variation signals can be detected from the light curve of BS\_Li23, respectively. Meanwhile, the `PHOT\_VARIABLE\_FLAG' of this star is `NOT\_AVAILABLE' in {\it Gaia} DR3.
Based on existing results, we believe that BS\_Li23 has a high probability of being a BSS single star produced by binary mergers. 

In the theoretical hypothesis of mass transfer in binary stars to form BSS, the orbital angular momentum of binary stars is accompanied by material transferred to the BSS, resulting in the newly formed BSS having a higher rotational speed \citep{2002MNRAS.332...49S, 2013ApJ...764..166D, 2017A&A...606A.137M}. Subsequently, due to the influence of the magnetic field, the BSS undergoes spin-down. Regarding observation, \cite{2023NatCo..14.2584F} confirmed 91 (about 28\%) fast-rotating BSSs through high-resolution spectral observations of 320 BSSs. Furthermore, \cite{2018ApJ...869L..29L} studied the relationship between the BSS rotational velocity and age in open clusters. They also proposed that the BSS rotation begins to slow down in several hundred million year timescales, with a spin-down timescale of approximately 1-2~Gyr. 

\begin{figure}
   \centering
   \includegraphics[width=0.5\textwidth]{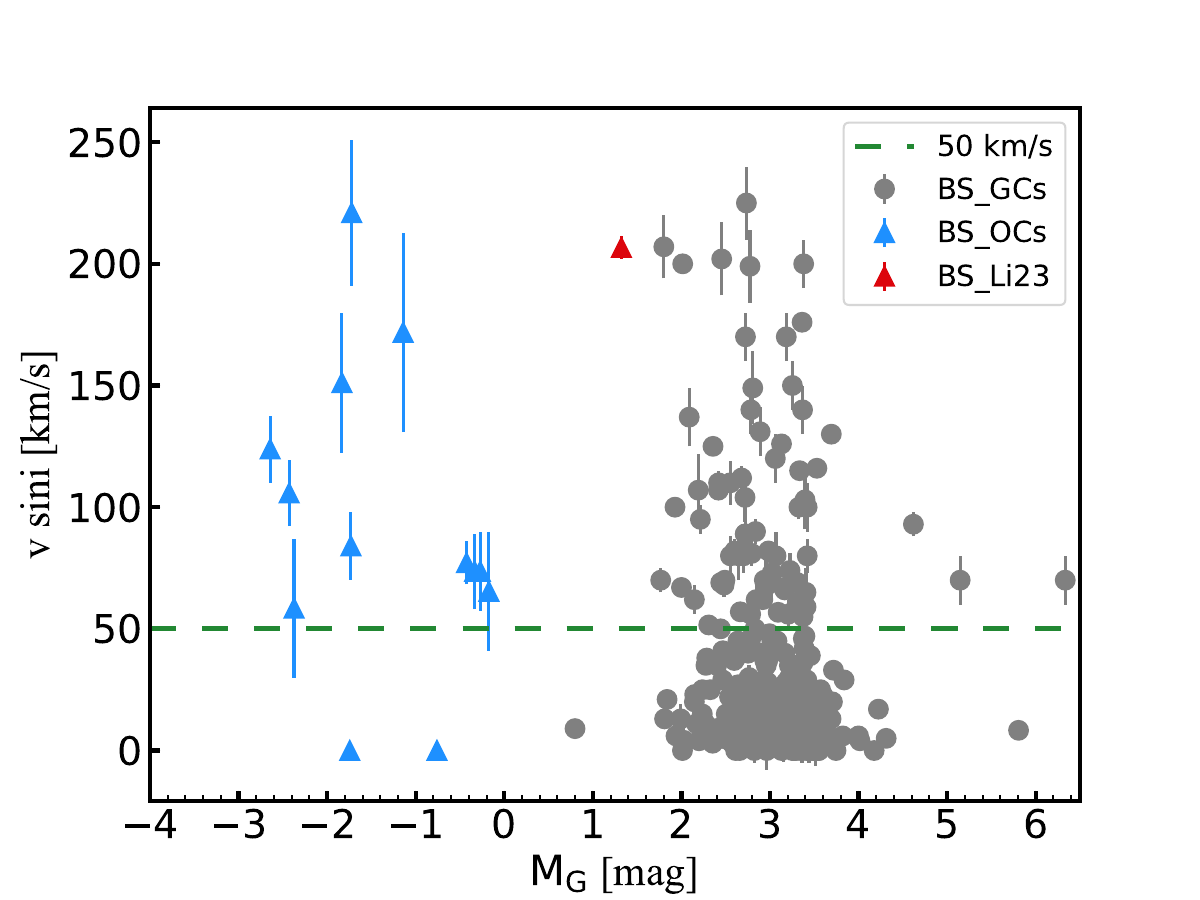}
      \caption{Rotational velocity distributions as a function of absolute G magnitude. the BSSs in OCs reported by \cite{2023A&A...672A..81L} are plotted as triangles, while BS\_Li23 is marked as a red triangle. BSSs in GCs from \citet{2023NatCo..14.2584F} are denoted by grey dots. The $v\ \mathrm{sin}i$ errors for blue triangles, red triangles, and grey dots are from {\it Gaia} DR3, Sect. \ref{sect:spect} of this work, and \cite{2023NatCo..14.2584F}, respectively. Distance module and reddening for BSSs in GCs are all from website \url{https://archive.stsci.edu/missions-and-data/tess}, while those for BSSs in OCs are from \cite{2023A&A...672A..81L}.}
         \label{fig: paper2_BS_vsini}
\end{figure}

Based on the spectral fitting results of LAMOST, the projected rotational velocity of BS\_Li23 is $206.9\pm4.9$~km $\rm s^{-1}$. We then compared BS\_Li23 with the 320 BSSs in GCs \citep{2023NatCo..14.2584F} and 13 BSSs in OCs \citep{2023A&A...672A..81L}. 
As shown in Fig. \ref{fig: paper2_BS_vsini}, it can be seen that BS\_ Li23 exhibits a characteristic of higher rotational velocity than most BSSs in both globular and open clusters observed so far which implies that BS\_Li23 may have not spun down yet.
The rapid rotational velocity of a BSS may indicate its recent formation \citep{2019ApJ...881...47L}. Based on the rough gyro-age relationship in \cite{2018ApJ...869L..29L}, we expect the age of BS\_Li23 to be restricted within a few hundred million years. Incidentally, NGC 752, the host open cluster of BS\_Li23, is almost the oldest one among these OCs. Therefore, BS\_Li23 appears to be more similar to the absolute magnitude of BSSs in globular clusters.

\section{Reevaluation of a formerly labeled BSS \label{sect:D94}}
Before {\it Gaia} era, the comprehensive member catalogs of NGC 752 were from \cite{1994PASP..106..281D}, \cite{1998A&A...339..423M}, and \cite{2018ApJ...862...33A}, while the determination of stellar members of NGC 752 is mainly based on ground observed proper motions and/or radial velocities.
In \cite{1994PASP..106..281D}, a star (BS\_D94, {\it Gaia} DR3 source ID 342856660737416320) was identified as a probable member and located within the defined region of BSSs in the CMD (see Fig. 3 in \cite{1994PASP..106..281D}). 
This star had retained as a cluster member of NGC 752 in subsequent studies \citep{1998A&A...339..423M, 2018ApJ...862...33A}. Until the {\it Gaia} mission provided higher-precision astrometric data, BS\_D94 has been excluded from the members of NGC 752 in most updated OC catalogs \citep{Cantat-Gaudin_2020, 2020A&A...640A...1C, 2021MNRAS.504..356D}. \cite{2021MNRAS.505.1607B} and \cite{2022MNRAS.514.3579B} expanded the search area for NGC 752 member stars. BS\_D94 has still not been identified as the cluster member in their works.

BS\_D94 was not included by \cite{2023A&A...672A..81L} in the initial sample for cluster membership determination. To investigate whether the star is a cluster member of NGC 752, we compared the characteristics of BS\_D94 with that of cluster members ($P_{m} \geq 0.9$) identified by \citep{2023A&A...672A..81L}. In Fig. \ref{fig: paper2_members4d}, the proper motion and parallax of BS\_D94 are considerably different from those high-probability member stars. The radial velocity of BS\_D94 from {\it Gaia} DR3 is $-3.81$~km $\rm s^{-1}$, which also significantly deviates from the average radial velocity value of NGC 752. We suggest that BS\_D94 is likely a background field star that only happened to project near the cluster's center.

\section{Summary\label{sect:sum}}
In the census work of BSSs by \cite{2023A&A...672A..81L}, a cluster member in the open cluster NGC 752 tidal tails was identified and classified as a BSS. We named this star BS\_Li23 in this work and performed a detailed investigation.

We systematically collected and analyzed all available spectral and photometric data of BS\_Li23. Firstly, the projected rotational velocity of BS\_Li23 was derived as 206.9~km $\rm s^{-1}$ based on a LAMOST DR9 low-resolution spectrum with measurement precision of about 10~km $\rm s^{-1}$.
Subsequently, the mass of BS\_Li23 was estimated by fitting its SED, combined from all available photometric datasets. The calculated mass of BS\_Li23 is $1.859^{+3.62}_{-0.936} M_{\odot}$, derived from the best-fitting result of stellar gravity and radius.
The SED of BS\_Li23 can be appropriately represented by a single-star model, also compatible with the analysis of time series photometric data from {\it TESS}. Finally, we provided the fundamental parameters, estimated projected rotational velocity, and stellar mass of BS\_Li23 in Table \ref{table:1}.

Furthermore, we discussed the possible formation pathway and age of BS\_Li23. Considering that this star was discovered in a low-density environment of tidal tail in the open cluster NGC 752, We believe that the stellar collision process could not practically be the formation pathway of this BSS. Then, because of the lack of FUV excess and flux variation, we speculate that this BSS was formed through binary evolution and has now merged into a single star. According to the formation theory of BSSs and the spin-down mechanism, the fast-rotating characteristic of BS\_Li23 further demonstrates that it is a recently formed BSS.

It is worth noting that a star (BS\_D94) was previously classified as a BSS in the literature. However, using the {\it Gaia} accurate astrometric data, we found that BS\_D94 is significantly different in its kinematic and parallax parameters compared to high-probability member stars ($P_{m} \geq 0.9$) identified by \cite{2023A&A...672A..81L}. We believe it is a background field star but coincidentally projected into the central region of NGC752.

In the {\it Gaia} era, our understanding of the spatial scale of open clusters has greatly expanded previous knowledge \citep{2019A&A...624A..34Z,2019A&A...621L...3M,2022RAA....22e5022B}. We now know that open clusters have not only high-density core components but also low-density extended outer halo components, which can even extend over a hundred pc \citep{2022AJ....164...54Z,2023ApJS..265...12Q}. Expanding the search region to hunt out more BSSs in the open clusters is essential. Moreover, exploring the BSS in the external areas of OCs will provide new insight to investigate the different formation mechanisms of BSSs and the impact of varying cluster environments on BSS formation \citep{2012Natur.492..393F, 2023ApJ...950..145F}.

\begin{acknowledgements}
We express our gratitude to the anonymous referee for their valuable comments and suggestions, which are very helpful in improving our manuscript.
This work is supported by the National Natural Science Foundation of China (NSFC) through grants 12090040, 12090042, and 12073060. 
J. Z. would like to acknowledge the Youth Innovation Promotion Association CAS, the science research grants from the China Manned Space Project with NO. CMS-CSST-2021-A08, the Science and Technology Commission of Shanghai Municipality (Grant No. 22dz1202400).
D. J. would like to acknowledge the National Key R\&D Program of China No. 2021YFA1600403, the National Natural Science Foundation of China (Nos. 12073070, 12333008, 12288102, 12090040/3), International Centre of Supernovae, Yunnan Key Laboratory (No. 202302AN360001), Yunnan Revitalization Talent Support Program -- Science \& Technology Champion Project (NO. 202305AB350003), the Yunnan Ten Thousand Talents Plan Young \& Elite Talents Project, and Yunnan Fundamental Research Project (No. 202201BC070003, 202401BC070007).
S. Q. acknowledges the financial support provided by the China Scholarship Council program(Grant No. 202304910547).

This work has made use of data from the European Space Agency (ESA) mission {\it Gaia} (\url{https://www.cosmos.esa.int/gaia}), processed by the {\it Gaia} Data Processing and Analysis Consortium (DPAC,\url{https://www.cosmos.esa.int/web/gaia/dpac/consortium}). Funding for the DPAC has been provided by national institutions, in particular, the institutions participating in the {\it Gaia} Multilateral Agreement.

Guoshoujing Telescope (the Large Sky Area Multi-Object Fiber Spectroscopic Telescope LAMOST) is a major national scientific Project built by the Chinese Academy of Sciences. Funding for the project has been provided by the National Development and Reform Commission. LAMOST is operated and managed by the National Astronomical Observatories, Chinese Academy of Sciences.

This paper includes data collected with the TESS mission, obtained from the MAST data archive at the Space Telescope Science Institute (STScI). Funding for the TESS mission is provided by the NASA Explorer Program. STScI is operated by the Association of Universities for Research in Astronomy, Inc., under NASA contract NAS 5–26555.
\end{acknowledgements}

\bibliographystyle{aa} 
\bibliography{ref} 

\end{CJK}
\end{document}